\documentclass[aps,pra,twocolumn,longbibliography,superscriptaddress,floatfix,10pt]{revtex4-2}
\usepackage{amsfonts,amssymb,dsfont}
\usepackage[sumlimits,intlimits]{amsmath}
\usepackage{graphicx}
\usepackage{color}
\usepackage{mathrsfs}
\usepackage{textcomp}
\usepackage{verbatim}
\usepackage{xr-hyper}
\usepackage[colorlinks=true,citecolor=blue]{hyperref}
\usepackage{bm}
\usepackage{braket}
\usepackage{qcircuit}
\usepackage{times}
\usepackage{booktabs}
\usepackage{soul}
\usepackage{tikz}
\usepackage[acronym]{glossaries}
\glsdisablehyper 
\usepackage{MnSymbol}
\usepackage{tikz} 
\usepackage{mathtools}

\usepackage[capitalise,compress]{cleveref}
\crefrangelabelformat{equation}{\textup{(#3#1#4)}--\textup{(#5#2#6)}}

\newcommand{\dket}[1]{\left|{#1}\right\rrangle}
\newcommand{\dbra}[1]{\left\llangle {#1}\right|}
\newcommand{\dbraket}[1]{\left\llangle {#1}\right\rrangle}
\newcommand{\tr}[1]{{\rm Tr}\left[#1\right]}

\newacronym{aces}{ACES}{averaged circuit eigenvalue sampling}
\newacronym{qcvv}{QCVV}{quantum characterization, verification, and validation}
\newacronym{gst}{GST}{gate-set tomography}
\newacronym{rb}{RB}{randomized benchmarking}
\newacronym{irb}{IRB}{interleaved randomized benchmarking}
\newacronym{cz}{CZ}{controlled-Z}
\newacronym{qec}{QEC}{quantum error correction}
\newacronym{ptm}{PTM}{Pauli transfer matrix}
\newacronym{cb}{CB}{cycle benchmarking}
\newacronym{cptp}{CPTP}{completely positive and trace preserving}
\newacronym{tc}{TC}{tunable coupler}
\newacronym{povm}{POVM}{positive operator valued measure}
\newacronym{mem}{MEM}{measurement error mitigation}
\newacronym{spam}{SPAM}{state-preparation and measurement}

\begin{document}

\title{Characterizing Superconducting Qubits using Averaged Circuit Eigenvalue Sampling}

\author{Tauno Palomaki}
\affiliation{Johns Hopkins Applied Physics Laboratory, Laurel, Maryland 20723, USA}

\author{Shu Xin Wu}
\affiliation{Johns Hopkins Applied Physics Laboratory, Laurel, Maryland 20723, USA}

\author{Noah Huffman}
\affiliation{Johns Hopkins Applied Physics Laboratory, Laurel, Maryland 20723, USA}
\affiliation{Stanford University, Palo Alto, CA, 94306, USA}

\author{Samuel D. Park}
\affiliation{Johns Hopkins Applied Physics Laboratory, Laurel, Maryland 20723, USA}

\author{James Shackford}
\affiliation{Johns Hopkins Applied Physics Laboratory, Laurel, Maryland 20723, USA}

\author{Ben DalFavero}
\affiliation{Johns Hopkins Applied Physics Laboratory, Laurel, Maryland 20723, USA}
\affiliation{Michigan State University, East Lansing, Michigan 48824, USA}

\author{Leigh Norris}
\affiliation{Johns Hopkins Applied Physics Laboratory, Laurel, Maryland 20723, USA}

\author{Ryan Sitler}
\affiliation{Johns Hopkins Applied Physics Laboratory, Laurel, Maryland 20723, USA}

\author{Paraj Titum}
\affiliation{Johns Hopkins Applied Physics Laboratory, Laurel, Maryland 20723, USA}
\affiliation{Johns Hopkins University, Baltimore, Maryland 21218, USA}

\author{Kevin Schultz}
\affiliation{Johns Hopkins Applied Physics Laboratory, Laurel, Maryland 20723, USA}

\begin{abstract}
Efficient characterization of noise during quantum gate operations is an essential step to building and scaling up a quantum computer.  One such protocol is \acrfull{aces} which efficiently characterizes a noisy gate set by reconstructing a Pauli noise model for a each gate. Here we utilize the \acrshort{aces} protocol to characterize two coupled superconducting qubits. For accurate reconstruction, we tailor the noise via Pauli twirling and account for measurement errors. We verify the accuracy of the protocol by comparing the predicted gate fidelities to that extracted from conventional benchmarking approaches, such as interleaved randomized benchmarking. Furthermore, we demonstrate the efficacy of \acrshort{aces} in accurately identifying specific noise sources by reconstructing injected phase errors in the two-qubit gates.
\end{abstract}
\maketitle
\section{Introduction}

Techniques for efficient estimation of noise occuring during quantum operations is a cornerstone for characterizing and improving the performance of gates in quantum computing. A wide variety of \acrfull{qcvv} protocols have been developed to quantify these errors, and they can be categorized based on the amount of information about the noise processes being extracted. Some protocols estimate a single metric~(e.g., average gate fidelity using \acrfull{irb}~\cite{Magesan_2012}), and others might reconstruct the full quantum channel representing a gate or a set of gates~(e.g., \acrfull{gst}~\cite{Nielsen2021gatesettomography}); see Ref.~\onlinecite{2024_Hashim} for a pedagogical review. However, with quantum computers scaling up in size to hundreds of qubits, these approaches of benchmarking become impractical. For example, tomographic approaches such as \acrshort{gst} require exponentially many measurements.  Even for \acrshort{qcvv} protocols such as \acrfull{rb}, where naively one might expect favorable scaling with the number of qubits $n$, the protocol becomes impractical as the implementation of a single $n$-qubit Clifford gate requires $O(n^{2}/\log{n})$ elementary gates. Thus, it has become necessary to develop efficient \acrshort{qcvv} protocols that extract error model that can be used for predictive modeling of these noisy quantum devices.

It is possible to dramatically reduce the required number of measurements needed for benchmarking by making a couple of assumptions about the relevant error processes. First, we restrict the errors to Pauli channels only, i.e., a Pauli noise model. An $n$-qubit Pauli channel can be viewed as an error model that probabilistically applies an $n$-qubit Pauli error and thus, is described by $4^n-1$ parameters (as opposed to $16^n-1$ parameters for the most general process matrix for an error channel). Second, one can further reduce the number of parameters in the error model with the assumption of locality of errors. For example, if the quantum circuit is composed of only single or two-qubit operations, one might opt to restrict the error processes to one or two qubits and be described by at most $4^2-1=15$ parameters per operation. Under these assumptions, several approaches have been developed to characterize qubits that broadly fall under the umbrella of Pauli noise learning~\cite{Erhard2019,Flammia2020,Harper2020,Huang2022,2023_Chen, 2023_Hashim,Hines2024,Berg2024,ChenPRL2024,2022_Flammia,2024_Pelaez, 2025_Chen, 2025B_Hocking,2025A_Hockings, 2025_Iyer}.  One of the protocols is \acrfull{aces}~\cite{2022_Flammia, 2024_Pelaez, 2025A_Hockings, 2025B_Hocking}. \acrshort{aces} identifies a minimal set of  Clifford circuits with suitable Pauli initial states and measurements that can be used to fully reconstruct the parameters of an assumed error model. This protocol provides a flexible framework to extract the noise parameters from an arbitrary set of Clifford quantum circuits. For example, prior work has utilized \acrshort{aces} for gate error estimation using random \acrfull{cb} circuits~\cite{2022_Flammia} and \acrfull{qec} circuits~\cite{2025B_Hocking}. 

While the assumption of a Pauli noise model might seem restrictive, many common errors in quantum computing are well described by it, for example, dephasing or depolarizing errors. However, certain physically-relevant noise processes measured in qubit devices are not intrinsically Pauli. Examples of such processes include spontaneous relaxation in qubits ($T_1$ error) which can be modeled by an amplitude damping channel and coherent gate errors. Fortunately, such errors can be tailored into a Pauli channel through a procedure known as Pauli twirling. Pauli twirling replaces the desired quantum operation to be executed by one that is conjugated by random gates chosen from the Pauli group, where the gate is randomized every shot. It is straightforward to show that the desired noise-tailoring is achieved when the quantum operation is averaged over the Pauli group; however, this proof assumes that the error process is Markovian, i.e., it is a \acrfull{cptp} map and is uncorrelated in time. Twirling has been utilized in a wide variety of context including benchmarking~\cite{Magesan2011}, compilation~\cite{Hashim2021}, and error mitigation~\cite{Temme2017, Ware2021}.

In this paper, we experimentally use \acrshort{aces} to characterize the error processes in two coupled transmon qubits. Our intention is to investigate the efficacy of \acrshort{aces}  as a characterization protocol in comparison to standard approaches.  The device consists of  two fixed-frequency qubits that are coupled by a tunable coupler. This tunable coupler enables us to perform fast entangling CZ gates~\cite{2021_Sung, 2021_Stehlik, 2020_Xu,2020_Collodo}. As the one-qubit gates have much higher fidelities than the two-qubit gate and measurement operations, we Pauli twirl both the two-qubit gate and measurement operations to tailor their noise into Pauli channels. A Pauli error model for all the gates in the gate set is estimated using a gradient-descent based optimization procedure.  We find that the predicted gate fidelities as extracted from the protocol are consistent with that extracted from \acrshort{rb}-based approaches. Since \acrshort{aces} extracts the Pauli channel corresponding to a gate error, it provides much more detailed information about the nature of the errors beyond just the average fidelity. In order to test the accuracy of the reconstruction from \acrshort{aces} and tailoring noise through Pauli twirling, we inject intentional coherent errors following the application of CZ gates. We demonstrate that \acrshort{aces} accurately estimates the magnitude of this intentional error as an added contribution to the estimated Pauli error process. In conclusion, our results indicate that \acrshort{aces} provides a reliable framework to simultaneously characterize all operations on a quantum device. 

This paper is organized as follows. We start with a concise introduction of the \acrshort{aces} protocol in \cref{sec:background}. In \cref{sec:methods}, we provide details of the experimental implementation of the protocol including the details of the device. \cref{sec:results} summarizes our results from analyzing the experimental data obtained from the device. Finally, we conclude with a discussion of possible future directions in \cref{sec:discussion}.

\section{Background}\label{sec:background}
In this section, we provide a concise overview of the Pauli group, Pauli channels, and twirling before summarizing the \acrshort{aces} protocol for estimating Pauli error rates of a family of gates from experimentally obtained data.

\subsection{Preliminaries}

\noindent{\it Pauli Group}--\ 
The one qubit Pauli group is the group generated by the Pauli matrices, i.e., $\tilde{\mathcal{P}}_1=\langle X, Y,Z\rangle$. The $n$-qubit Pauli group, is defined by the $n$-fold tensor products of the elements of the one qubit Pauli group. The elements of the Pauli group are specified by an $n$-qubit Pauli string and an overall phase in $\{\pm 1,\pm i\}$. Since the overall phase is typically not relevant for measured probabilities, we will restrict ourselves to the quotient group that sets the overall phase set to $+1$,
\begin{align}
    \mathcal{P}_1&:=\tilde{\mathcal{P}}_1/\braket{\pm iI}\equiv\left\{I,X,Y,Z\right\},\\
    \mathcal{P}_n&:=\{P_1\otimes\cdots \otimes P_n|P_1,\cdots P_n \in \mathcal{P}_1\}.
\end{align}
We will refer to $\mathcal{P}_n$ as the Pauli group as defined above with dimension $d=4^{n}$. The elements of the Pauli group are Hermitian.

\noindent{\it Pauli Basis Representation}--\ 
Multi-qubit quantum states can be represented using the Pauli group as a basis. Formally, the quantum state space is spanned by the Pauli basis, $\{\dket{P_a}\}$, where the inner-product is defined by the Hilbert-Schmidt operator $\dbraket{A|B}=\frac{1}{\sqrt{d}}\tr{A^\dagger B}$. The Pauli group elements, $\dket{P_a}$, form an orthonormal basis under this definition of the inner product, $\dbraket{P_a|P_b}=\delta_{ab}$. Thus, an arbitrary $n$-qubit density matrix can be expanded as,
\begin{align}
    \dket{\rho}=\sum_a \rho_{a} \dket{P_a},
\end{align}
with $\rho_a=\dbraket{P_a|\rho}$ and $P_a\in\mathcal{P}_n$; an arbitrary quantum state is represented as a $d$-dimensional vector. 

Next, we consider the representation of a general quantum operation using the Pauli basis, $\mathcal{E}:\rho\rightarrow\mathcal{E}(\rho)$, also known as the \acrfull{ptm}. Consider the case of a channel specified by a set of Kraus operators $\{K_j\}$, $\mathcal{E}(A)=\sum_jK_j A K_j^\dagger$. The \acrshort{ptm} representation for the channel is given by the matrix $R_\mathcal{E}$ that defines the action of the channel in the Pauli basis as,
\begin{align}
    R_{\mathcal{E}}&=\sum_{a,b=1}^d\left[R_{\mathcal{E}}\right]_{ab} \dket{P_a}\dbra{P_b}\\
    \left[R_{\mathcal{E}}\right]_{ab}&=\dbraket{P_a|\mathcal{E}(P_b)}=\frac{1}{\sqrt{d}}\sum_j\tr{P_aK_jP_bK_j^\dagger}\\      
\end{align}
The \acrshort{ptm}, $R_\mathcal{E}$ is a $d\times d$ matrix. The output state of a quantum channel is simply obtained through matrix multiplication in the Pauli basis, $\dbraket{P_a|\mathcal{E}(\rho)}=\sum_b \left[R_{\mathcal{E}}\right]_{ab}\rho_b$.

\noindent{\it Pauli Channel}--\ 
Pauli channels are quantum channels that have a diagonal \acrshort{ptm},
\begin{equation}
    R_{\mathcal{E}}=\sum_{a=1}^d \lambda_a \dket{P_a}\dbra{P_a}
\end{equation}
where, $\lambda_a$ are known as the Pauli eigenvalues since the Pauli states, $\dket{P_a}$, are eigenstates of the channel. It is straightforward to show that the Kraus operators for the channel are  of the form $\sqrt{p_a}P_a$, where, $p_a$ are called Pauli error rates. Intuitively, a Pauli channel can be viewed as a stochastic application of a Pauli error $P_a$ with a probability set by the Pauli error rate. The Pauli error rates are related to the Pauli eigenvalues by a Hadamard transform. As an example, we provide the transformation for one-qubit Pauli channel,
\begin{align}
    \left(\begin{matrix}
        \lambda_I\\
        \lambda_X\\
        \lambda_Y\\
        \lambda_Z
    \end{matrix}\right)=
    \underbrace{\left(
        \begin{matrix}
            1 & 1 & 1& 1\\
            1 & 1 & -1& -1\\
            1 & -1 & 1& -1\\
            1 & -1 & -1& 1
        \end{matrix}
    \right)}_{\mathbb{H}}
    \left(\begin{matrix}
        p_I\\
        p_X\\
        p_Y\\
        p_Z
    \end{matrix}\right)\label{eq:hadamard}
\end{align} 
where $p_a$ and $\lambda_a$ $a\in \{I,X,Y,Z\}$ are the single-qubit Pauli error rates and eigenvalues respectively, and $\mathbb{H}$ is the Hadamard transform matrix. Assuming no leakage, i.e., the channel is trace preserving, $\lambda_I=1$. This transformation can be straightforwardly generalized for Pauli channels acting on two or more qubits.

\noindent{\it Noisy Measurements}--\ 
We define ideal projective measurements by a \acrfull{povm}, $\{E_j\}$; $E_j\equiv\ket{j}\bra{j}$ where $j$ represents a bit-string output $j\in\{0,1\}^{\otimes n}$. The outcome probabilities for a state $\rho$ when measured by this \acrshort{povm} is $p_j=\sqrt{d}\dbraket{E_j|\rho}$. 
In experiments, measurement error can come from two distinct sources: (i) 'quantum' (e.g., relaxation of the state), and (ii) 'classical' (e.g., mis-labeling of outcome state). In practice, these two errors can be difficult to disambiguate, and the measurement error probability is specified by a single confusion matrix $C$, where the row specifies the true outcome and column the measured outcome, and $\mathcal{C}_{ij}=p(j|i)$. Experimentally, $\mathcal{C}$ is estimated by preparing and measuring in the different basis states $\ket{j}$.  The noisy outcome probabilities, $\tilde{p}_i$ can be related to the ideal outcome probabilities $p_{j}$,
\begin{align}
    \tilde{p}_i=\sum_j\mathcal{C}_{ji}p_j
\end{align} 
In general, the confusion matrix is a real matrix with each row normalized to 1.

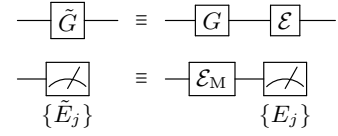
\begin{figure}
    \centering
    \begin{tabular}{c}
         \Qcircuit @C=1em @R=0em {
           & \gate{\tilde{G}} & \qw {\rule{2em}{0em}  \equiv \rule{0em}{0em}} & & & \gate{G}&\gate{\mathcal{E}}& \qw \\
           & \push{\rule{0pt}{1em}} \\ 
           &\meter &  {\rule{2em}{0em}  \equiv \rule{0em}{0em}} & & & \gate{\mathcal{E}_{\rm M}}& \meter\\
           &\push{\rule{0pt}{1.1em}{\{\tilde{E}_j\}}}&&&&&\push{\rule{0pt}{1.1em}{\{{E}_j\}}}
        } 
    \end{tabular}
    \caption{Error model for a noisy Clifford gate~(G), and noisy measurement ($\{\tilde{E}_j\}$). We model gate errors as occurring \emph{after} the gate, and measurement errors \emph{before} the gate.}
    \label{fig:error-model}
\end{figure}

For simulating noisy measurements, disregarding the post-measured state, the noisy output probabilities can be modeled as arising from an error channel~($\mathcal{E}_M$) that precedes an ideal projective measurement (see \cref{fig:error-model}),
\begin{align}
    \tilde{p}_i=\sqrt{d}\dbraket{\tilde{E}_i|\rho}=\sum_j\sqrt{d}\dbraket{E_i|R_{\mathcal{E}_M}|E_j}p_j,
\end{align} 
where $\{\tilde{E}_j\}$ represents the noisy \acrshort{povm}. Under this model, the confusion matrix is given by, $\mathcal{C}_{ji}=\sqrt{d}\dbraket{\tilde{E}_i|E_j}$. We note a couple of observations that are relevant to this work. First, when $\mathcal{E}_{\rm M}=\sum_a \lambda_{{\rm M},a} \dket{P_a}\dbra{P_a}$ is a Pauli channel, it is straightforward to show that the confusion matrix $\mathcal{C}_{ji}$ must be symmetric. Second, an asymmetric confusion matrix can be obtained by considering a non-Pauli $\mathcal{E}_{\rm M}$; e.g., a single-qubit generalized amplitude damping channel with probability of damping $\gamma$, and steady state ground state probability $\beta$. Explicitly, the \acrshort{ptm} for this damping channel is,
\begin{align}
    R_{\mathcal{E}_M}=\left(
        \begin{matrix}
            1 & 0 & 0& 0\\
            0 & \sqrt{1-\gamma} & 0& 0\\
            0 & 0 & \sqrt{1-\gamma}& 0\\
            \gamma(2\beta-1) & 0 & 0& 1-\gamma
        \end{matrix}
    \right),
\end{align}
which recovers the confusion matrix,
\begin{align}
    \mathcal{C}=\left(\begin{matrix}
        1+\gamma(\beta-1)& \gamma(1 -\beta) \\
        \gamma \beta& 1-\gamma\beta
    \end{matrix}\right).
\end{align}
The parameters $\gamma$ and $\beta$ can be chosen to account for any asymmetry in $\mathcal{C}$.

\noindent{\it Pauli Twirling for Gates}--\ 
Twirling refers to the procedure of conjugating a channel by gates chosen uniformly randomly from a set; Pauli twirling corresponds to uniformly sampling the Pauli group. Pauli twirling converts an arbitrary \acrshort{cptp} quantum channel into a Pauli channel. This can be straightforwardly derived by observing that the \acrshort{ptm} representation of a Pauli gate is specified by the Hadamard transform matrix, $R_{P_a}=\sum_b\mathbb{H}_{ba} \dket{P_b}\dbra{P_b}$. Thus, the Pauli-twirled channel, $R^P_{\mathcal{E}}$ is given by,
\begin{align}
    R^P_\mathcal{E}&= \frac{1}{d}\sum_{a=1}^d R_{P_a^\dagger} R_\mathcal{E} R_{P_a}\\
    &= \frac{1}{d}\sum_{a,b,c=1}^d \mathbb{H}_{ca} \mathbb{H}_{ba} \dket{P_c}\dbra{P_c} R_\mathcal{E} \dket{P_b}\dbra{P_b}\\
    &=\sum_{c=1}^d\dbra{P_c} R_\mathcal{E} \dket{P_c}\dket{P_c}\dbra{P_c}
\end{align}
where, we utilized the orthogonality of the columns of the Hadamard matrix, $\sum_a \mathbb{H}_{ca}\mathbb{H}_{ba}=d\delta_{bc}$ in the final step.

In noisy implementation of quantum gates, the intended goal is to convert the noise channel into a Pauli channel, without changing the ideal gate operation. This can be achieved for Clifford gates using a modified Pauli twirling~(also referred to as the $G$-twisted twirl~\cite{2022_Flammia}). Consider an ideal Clifford gate $G$, and its noisy implementation, $\tilde{G}$ given in the \acrshort{ptm} representation, $R_{\tilde{G}}=R_\mathcal{E}R_G$, where $\mathcal{E}$ is the error process; see \cref{fig:error-model}. Since the Clifford group normalizes  the Pauli group, we can define $G^\dagger P_a G =P_{a^\prime}$. This identity can be equivalently written in the \acrshort{ptm} representation as $R_{G^\dagger} R_{P_a}R_{G}=R_{P_{a^\prime}}$. The $G$-twisted twirl is defined as,
\begin{align}
    R_{\tilde{G}}^P &= \frac{1}{d} \sum_{a=1}^d R_{P_{a}}R_{\tilde{G}}R_{P_{a^\prime}},\\
    &=\left(\frac{1}{d}\sum_{a=1}^d R_{P_a}R_{\mathcal{E}}R_{P_a}\right)R_G=R^P_\mathcal{E}R_G,
\end{align}
which achieves the desired twirl of the error in the gate. For a circuit, that is composed of several gates, the desired twirl of the circuit can be obtained by applying a random Pauli before each gate and the gate-conjugated Pauli after the gate.

\noindent{\it Pauli Twirling for Measurements}--\ 
Pauli twirling for measurements is a procedure to tailor the measurement error. The twirled error channel for measurements is defined as,
\begin{align}
    R^P_{\mathcal{E}_{\rm M}}=\sum_a \lambda_{{\rm M},a}\dket{P_a}\dbra{P_a},\ \textrm{and } \lambda_{{\rm M},a}=\dbraket{P_a|R_{\mathcal{E}_{\rm M}}|P_a}.\label{eq:measure-twirl-channel}
\end{align}
One qubit measurement twirling is achieved by randomly sampling the Pauli group $P_a \in \mathcal{P}_1$ and flipping the classical outcome when the Pauli chosen is $X$ or $Y$. By computing the elements of the confusion matrix, we can show that this procedure symmetrizes it and thus, effectively twirls the measurement error channel. Explicitly,
\begin{align}
    \mathcal{C}_{01}=&\frac{\sqrt{d}}{d}\big(\dbra{E_1} R_{\mathcal{E}_{\rm M}}R_I\dket{E_0}+\dbra{E_0} R_{\mathcal{E}_{\rm M}}R_X\dket{E_0}\nonumber\\
    &\qquad\dbra{E_0} R_{\mathcal{E}_{\rm M}}R_Y\dket{E_0}+\dbra{E_1} R_{\mathcal{E}_{\rm M}}R_Z\dket{E_0}\big)\\
    &=\sqrt{d}\dbra{E_1}R^P_{\mathcal{E}_{\rm M}}\dket{E_0},\\
    \mathcal{C}_{10}=&\sqrt{d}\dbraket{E_0|R^P_{\mathcal{E}_{\rm M}}|E_1}
\end{align}
where we have used $\dbra{E_{1,0}}R_{X,Y}=\dbra{E_{0,1}}$. Using the form of the twirled channel in \cref{eq:measure-twirl-channel}, it is clear that the confusion matrix becomes symmetric ($\mathcal{C}_{01}=\mathcal{C}_{10}$). In \cref{sec:methods}, we experimentally verify measurement twirling by examining the symmetry of the empirically obtained confusion matrix.

\subsection{Review of ACES}\label{sec:aces-review}
\acrshort{aces} is a characterization technique that simultaneously estimates Pauli error channels $\{\mathcal{E}_j\}$ associated with a collection of Clifford gates $\{G_j\}$ of interest and their noisy gates represented as $\{\tilde{G}_j=\mathcal{E}_jG_j\}$.~\cite{2022_Flammia}. In experiments, typically the noise is not expected to be Pauli, and thus, it is expected that all gates are appropriately twirled for the estimation procedure. The \acrshort{aces} procedure is built on three observations: 
\begin{enumerate}
    \item Pauli states, $\dket{P_a}$ satisfy a generalized eigenvalue equation for Clifford gates. For a Clifford gate $G_j$ and its noisy implementation, $\tilde{G}_j$, we have,
    \begin{align}
        R_G\dket{P_a}=\pm \dket{P_{a^\prime}}, \quad 
        R_{\tilde{G}_j}\dket{P_{a}}=\lambda_{j,a^\prime}\dket{P_{a^\prime}}
    \end{align}
    where $GP_{a}G^\dagger=P_{a^\prime}$, and the sign ($\pm$) for the noiseless gate can be efficiently computed classically and the noisy gate eigenvalue $\lambda_{j,a}$ is the eigenvalue for the Pauli noise channel associated with the noisy gate $R_{\mathcal{E}_j}\dket{P_a}=\lambda_{j,a}\dket{P_a}$.
    \item A circuit, $C$, composed of $T$ noisy Clifford gates also satisfies a generalized eigenvalue equation with eigenvalue $\Lambda_{C,a}$ which can be related back to the component gate eigenvalues,
    \begin{align}
        R_{C}\dket{P_{a}}= \Lambda_{C, a^\prime}\dket{P_{a^\prime}}, \quad 
        \Lambda_{C, a_{T+1}} = \prod_{k=1}^{T} \lambda_{k, a_{k+1}}, \label{eq:Aces-circuit-eigval}
    \end{align}
    where, $a\equiv a_1$, and $\lambda_{k,a_k}$ corresponds to the gate eigenvalue of the $k^{\rm th}$ gate in the circuit. Note, that the sign convention can be chosen such that all gate and circuit eigenvalues are positive.
  
  \item The circuit eigenvalue, $\Lambda_{C,a}$ can be efficiently estimated by sampling initial states that are prepared in the $\pm$ basis states $\{\ket{\psi^{\pm}_{a,b}}\}$ of the Pauli operator, $P_{a}\ket{\psi^{\pm}_{a,b}}=\pm\ket{\psi^{\pm}_{a,b}}$, where $b\in (1, ..., 2^{n-1})$ and $n$ is the number of qubits  ~\cite{Flammia2020}. Defining $\rho_{a,b}^{\pm}=\ket{\psi^{\pm}_{a,b}}\bra{\psi^{\pm}_{a,b}}$ , 
    \begin{align}
        \Lambda_{C,a^\prime}&=\dbraket{P_{a^\prime}|R_C|P_a}\\
        &=\sum_{b,b^\prime}\dbraket{\rho^+_{a^\prime,b^\prime}|R_C|\rho_{a,b}^{+}}+\dbraket{\rho^-_{a^\prime,b^\prime}|R_C|\rho_{a,b}^{-}}\nonumber\\
        &\ \ \ \  -\dbraket{\rho^+_{a^\prime,b^\prime}|R_C|\rho_{a,b}^{-}}-\dbraket{\rho^-_{a^\prime,b^\prime}|R_C|\rho_{a,b}^{+}}
    \end{align}
   In practice, we chose circuits such that $P_a$ and $P_a^\prime$ act non-trivially only on a small subset of qubits.
\end{enumerate}

\acrshort{aces} systematically identifies a complete set of unique circuit, and input and output Paulis, $\{(C_m,P_a,P_{a^\prime})\}$ that can be used to calculate the desired gate eigenvalues $\{\lambda_{j,a}\}$. Setting up the gate and circuit eigenvalues as a column vector and defining $\lambda_{\nu}=e^{-x_\nu}$ and $\Lambda_\mu = e^{-b_\mu}$, where $\nu\equiv(j,a)$ and $\mu\equiv(m,a)$, \cref{eq:Aces-circuit-eigval} can be recast as a linear system of equations,
\begin{align}
    \sum_{v} A_{\mu \nu} x_{\nu} = b_{\mu}, \label{eq:aces-eq}
\end{align}
where, each row of the design matrix, $A_{\mu \nu}$ corresponds to a unique tuple of circuit and input and output Paulis, and consists of integers depending on which gate eigenvalues appear in the circuit. If the error model has $M$ unknown gate eigenvalues, the \acrshort{aces} procedure can be setup to systematically search through different circuit and input Pauli combinations until a full rank design matrix is obtained. With a full rank design matrix one can invert the system of equations to obtain the gate eigenvalues and consequently the Pauli error rates.

\section{Methods}\label{sec:methods}

\subsection{Experimental Setup}\label{sec:expstep}

\begin{figure}
    \centering
    \makebox[\linewidth][c]{
    \begin{minipage}[b]{0.5\linewidth}
        \centering
        \begin{tikzpicture}
            \node[anchor=south west, inner sep=0] (image) at (0,0) {
                \includegraphics[angle=90,scale=0.14]{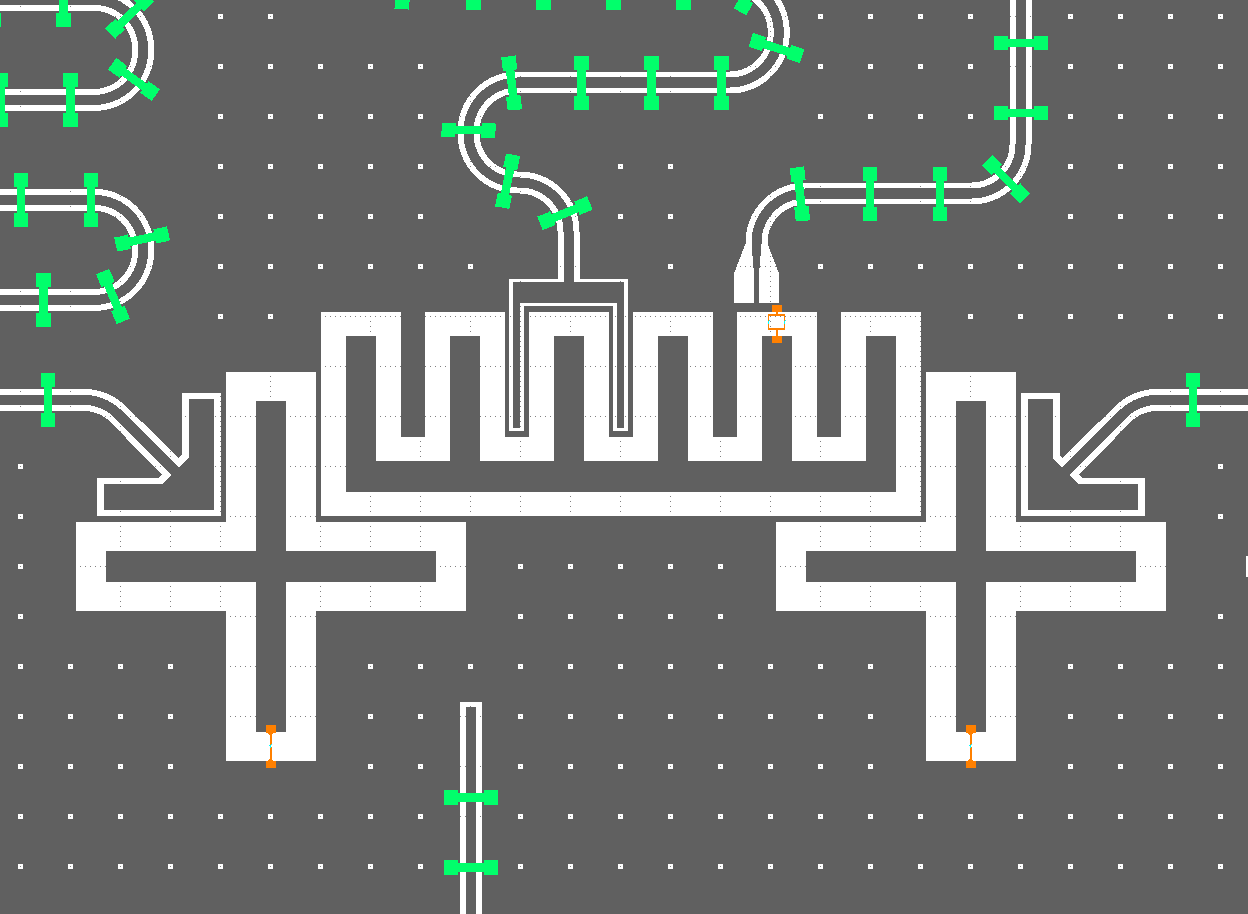}
            };
            \node[anchor=north west, font=\bfseries, text=white] at (image.north west) {(a)};
            \end{tikzpicture}
     \end{minipage}
    \begin{minipage}[b]{0.43\linewidth}
        \centering
        \begin{tikzpicture}
            \node[anchor=south west, inner sep=0] (tbl) at (0,0) {
                \begin{tabular}{|c|c|} 
                        \hline
                        Parameter &  Value \\
                        
                        \hline  
                        $f_{\rm Q0}$ &3.690 GHz \\    
                        $f_{\rm Q1}$ & 3.466 GHz \\
                        $f_{\rm TC}$ & 3.4 - 6.3 GHz \\
                        $\alpha_{\rm Q0}$ & 261 MHz \\
                        $\alpha_{\rm Q1}$ & 265 MHz \\
                        $\alpha_{\rm TC}$ & 95 MHz \\
                        $g_{\rm Q0-TC}$ & 49 MHz \\
                        $g_{\rm Q1-TC}$ & 46 MHz \\
                        $g_{\rm Q0-Q1}$ & $\sim$ 5 MHz \\

                        \hline
                \end{tabular}
            };
            \node[anchor=north west, font=\bfseries] at (-0.6,4.6) {(b)};
        \end{tikzpicture}
    \end{minipage}
    }
    \caption{Experimental details for the device used. (a)~Schematic of the superconducting qubit device consisting of two transmons coupled via a tunable coupler. (b) Estimated parameters of the device components. These parameters include frequency~($f$) and anharmonicity~($\alpha$) of the  two qubits ($\rm Q0$, $\rm Q1$) and the tunable coupler ($\rm TC$), and the transverse coupling ($g$) between each component pair.}\label{fig:DeviceSchematic}
\end{figure}

The quantum device utilized to demonstrate the \acrshort{aces} characterization procedure consists of two fixed-frequency transmons coupled with a \acrfull{tc}; see \cref{fig:DeviceSchematic}~(a) for a device schematic. The qubit frequencies are chosen to be separated enough to minimize cross-talk due to direct capacitive coupling between the qubits. The \acrshort{tc} is also capacitively coupled to the qubits which enables an effective $ZZ$ interaction between the qubits. We summarize the estimated qubit and coupler parameters in \cref{fig:DeviceSchematic}~(b). Qubit control and multiplexed readout is performed using a Quantum Machines OPX+.

The device is set up to perform universal quantum computation via the following standard set of quantum gates: $\mathcal{G}\equiv\{\sqrt{X_j},S_j,R_{z,j}(\phi), CZ, M_j\}$, with $j\in\{0,1\}$ indexing the two qubits. The one qubit $\sqrt{X}$ gates are performed using Gaussian pulses resonant with the qubit frequency. The phase gate ($S$) and the more general $R_z(\phi)$ gates are achieved through a virtual gate or frame rotation. The duration of all one qubit gates are fixed to $60$~ns; i.e., virtual gates are appended by an identity or delay gate to ensure that all single-qubit gates take the same total time. In order to minimize cross-talk during one qubit gate operations, the \acrshort{tc}-qubit frequency~($f_{\rm TC}$) is set $\sim2$~GHz above the qubit frequencies~($f_{\rm Q0}$ and $f_{\rm Q1}$) where the $ZZ$ interaction between the qubits is negligible ($\lesssim 10$~kHz). The $CZ$ gate is performed by flux-pulse-tuning the frequency of the \acrshort{tc} closer to resonance with $\rm Q1$, which results in an enhanced $ZZ$ interaction between the two qubits ~\cite{2021_Sung, 2021_Stehlik, 2020_Xu,2020_Collodo}. The  flux pulses used in this work are flattop Gaussian with a total duration between $250$ and $300$ ns. These pulses also change the individual qubit frequencies, the effects of which are removed by correction gates implemented via virtual frame rotations to each of the qubits. The measurement operations $M_j$ are performed via traditional dispersive readout of transmon qubits coupled to readout resonators.

We have benchmarked the single and two-qubit gates through \acrshort{rb} protocols. The one qubit average gate fidelities, as estimated from \acrshort{rb}, is $\sim99.8\%$. The fidelity of the $CZ$ gate is estimated via \acrshort{irb} and is $\sim98\%$. We discuss these traditional benchmarking results in greater detail in \cref{sec:results}.

\subsection{Pauli twirling implementation}\label{sec:twirling-implementation}

\begin{figure}[htb!]
    \centering
    \begin{tabular}{c}
         \Qcircuit @C=1em @R=0em {
           & \ctrl{2}  &\qw  & & & \gate{P_{a}} & \ctrl{2} & \qw & \gate{P_{c}}  & \qw   \\
           &     &       &         \push{\rule{0em}{0em}  \Rightarrow \rule{0em}{0em}}   \\
	 & \ctrl{-2} &\qw &  &  & \gate{P_{b}} & \ctrl{-2}& \gate{R_{z}(\Delta \theta)} &\gate{P_{d}} & \qw\gategroup{1}{6}{3}{9}{.7em}{--}\\
        & \push{\rule{0pt}{1em}} \\ 
            & \meter&  &       &           \Rightarrow \hspace{3.5em}  &\gate{P_a} &\meter&\cw\gategroup{5}{6}{5}{7}{.7em}{--}
        }
    \end{tabular}
    \caption{Composite gates incorporating Pauli twirling (represented by Pauli gates $P_{a}$) and coherent error injection (represented by $R_z$ gate). (Top) Twirled CZ gate with an optional injected coherent rotation error along the $z$-axis with the rotation angle set at  $\Delta \theta$. The Pauli gates $P_a$ and $P_b$ are randomly chosen for every CZ. (Bottom) Twirled measurement operation with a randomly chosen Pauli gate $P_a$, where the classical output bit is flipped when $P_a$ is $X$ or $Y$. }
    \label{fig:PauliTwirlError}
\end{figure}
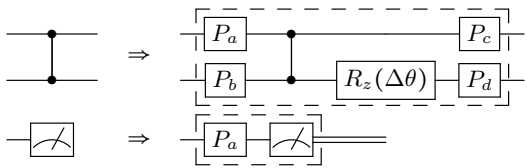

A necessary requirement for error estimation using \acrshort{aces} is that the errors are Pauli channels. As discussed in \cref{sec:background}, this can be achieved through Pauli twirling (G-twisted) which tailors arbitrary Markovian error processes during gate and measurement operations into Pauli channels. This procedure works when one qubit gate operations (i.e., the Pauli twirling gates) are noiseless. This scenario is essentially satisfied in the experimental implementation, where the dominant non-Pauli error processes occur during the two-qubit gate and single-qubit measurement operations; one qubit error rates are an order of magnitude lower than these errors. Thus, in our implementation of the  \acrshort{aces} protocol, we focus on twirling the errors occurring during these two operations. We define a composite $CZ$ gate and a composite measurement operation that each incorporate the Pauli twirling operation; this is shown schematically in \cref{fig:PauliTwirlError}. The schematic also includes an intentional rotation error $R_z(\Delta \theta)$ imediately after the bare CZ gate. As we discuss in later sections, \acrshort{aces} estimates the Pauli error probabilities associated with these composite operations.

Twirling is implemented in the experiments using the Quantum Machines OPX+ control hardware. Each $CZ$ gate in every experimental shot is subject to a new random twirl specified by a uniformly random two-qubit Pauli $(P_{a} \otimes P_{b})$ preceding the $CZ$ gate. Next, a look-up table is used to specify the corresponding 2-qubit Pauli $(P_{c} \otimes P_{d})=CZ (P_{a} \otimes P_{b}) CZ^\dagger $ that needs to be applied after the $CZ$ gate. As discussed in \cref{sec:background}, this (G-twisted) Pauli twirl tailors errors during the $CZ$ gate to Pauli channels. Twirling of errors during the measurement operations are done analogously. In this characterization procedure, we restrict to one qubit computational basis measurement operations at the end of quantum circuit (i.e., no mid-circuit measurements). Twirling of the measurement operations are achieved by randomly applying a Pauli gate prior to the measurement and flipping the measured outcome when the randomly chosen Pauli is either an $X$ or $Y$ gate~\cite{2023_Beale, 2025_Hashim}. As discussed in \cref{sec:background}, this twirls the measurement error into a Pauli channel.

\begin{figure}[htbt]
    \begin{minipage}[b]{\linewidth}
        \centering
        \begin{tikzpicture}
            \node (image) at (0.25,0) {
                \mbox{
                        \begin{tabular}{c}
                            \Qcircuit @C=1em @R=1em {
                                \lstick{\ket{0}} & \gate{X} & \ctrl{1} & \qw & \push{\ldots} &&\qw & \ctrl{1} & \qw&\qw& \\
                                \lstick{\ket{0}} & \gate{\sqrt{X}}      & \control \qw & \qw& \push{\ldots} &&\qw & \control \qw & \qw&\gate{\sqrt{X}} &\meter\\
                                &&&&&\hspace{-2em}\protect\underbrace{\hspace{7.5em}}_{\substack{\text{\small Odd number $(N_{\rm CZ}) $ of \vphantom{y}}\\ \text{\small composite CZ gates}}}\\
                                &
                            }               
                            \vspace{.5 cm}
                        \end{tabular}
                }
            };
            \node(image) at (0,-5) {
                \includegraphics[scale=.45]{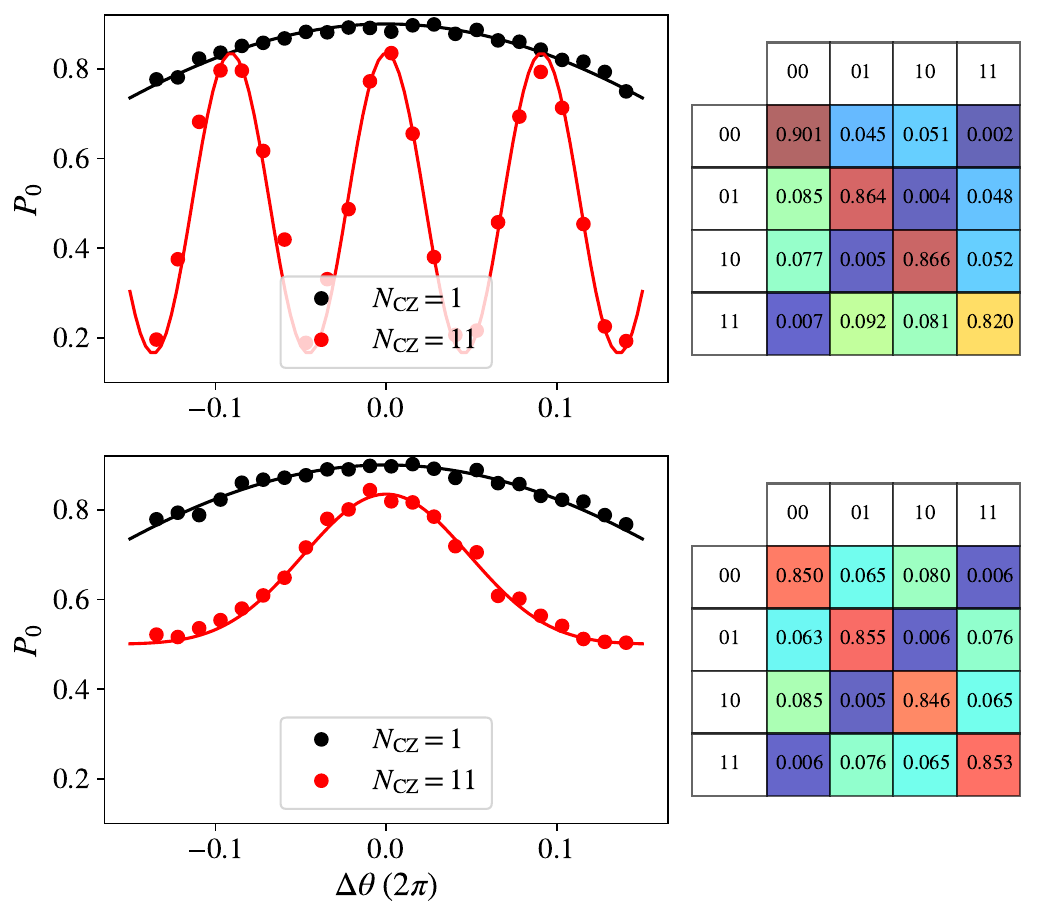}
            };
            \node[ font=\bfseries, text=black] at (-4,1.) {(a)};
            \node[ font=\bfseries, text=black] at (-4,-1.6) {(b)};
            \node[ font=\bfseries, text=black] at (1.4 ,-1.6) {(d)};
            \node[ font=\bfseries, text=black] at (-4,-5) {(c)};
            \node[ font=\bfseries, text=black] at (1.4,-5) {(e)};
        \end{tikzpicture}
    \end{minipage}
    \caption{Experimental demonstration of G-twisted twirled $CZ$ gates and measurement twirling. (a)~Circuit for benchmarking our twirled $CZ$. (b)~Measured probability for 1~(black) and 11~(red) repeated $CZ$-gates subject to an injected coherent error.  (c)~Same as (b), but G-twisted twirling each $CZ$-gate. (d) Experimentally measured confusion matrix without measurement twirling and (e) with measurement twirling, 10,000 shots per prepared state.}
  \label{fig:Twirled_Example}
\end{figure}
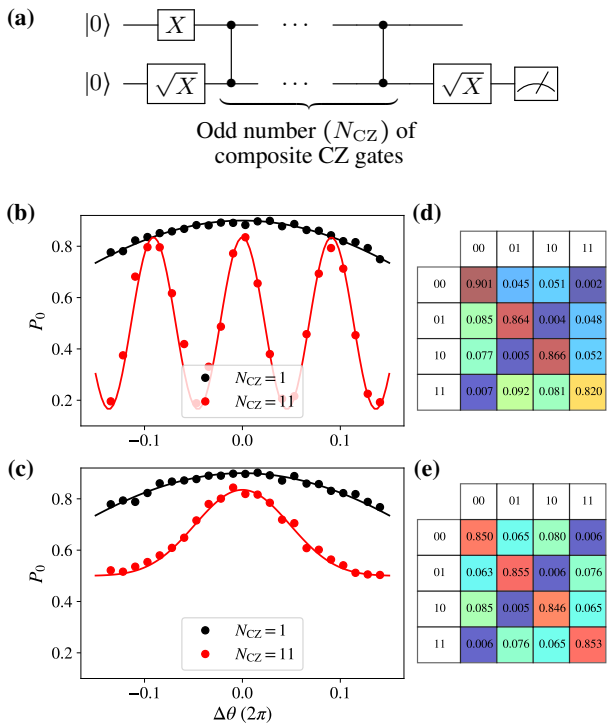

We test the efficacy of the twirling operation for tailoring gate errors using the circuit shown in \cref {fig:Twirled_Example} (a), where our CZ gate is now the composite CZ as described in  \cref{fig:PauliTwirlError} with an intentional coherent error $R_z(\Delta \theta)$. Such a coherent error should be tailored into a Pauli channel by the twirl. For an ideal sequence consisting of noiseless gates and measurement would yield the measured outcome of the second qubit, $P_{0}=1$. The outcome probability changes as a function of the magnitude of the injected coherent error, $\Delta \theta$, and the number of repeated CZ gates, $N_{\rm CZ}$. In the absence of twirling, where the Paulis in the composite $CZ$ gates are all set to identity gates, the outcome probability exhibits oscillations consistent with an $R_z(\Delta \theta)$ rotation on the second qubit. With twirling, these coherent oscillations are damped out, and the outcome probability decays in a manner consistent with a dissipative Pauli channel. We expect the following analytic forms for the survival probabilities for $N_{\rm CZ}$ gates,
\begin{align}
P^{\rm Coherent}_0(\Delta \theta)&=\frac{1}{2}(1+A_{N_{CZ}}\cos(N_{CZ}\Delta\theta))\\
P^{\rm Twirled}_0(\Delta\theta)&=\frac{1}{2}\left(1+A_{N_{CZ}}\cos^{N_{CZ}}(\Delta\theta)\right)
\end{align}
where, ideally, in the absence of any other error mechanisms $A_{N_{CZ}}=1$. The experimentally obtained survival probabilities are shown \cref{fig:Twirled_Example} in (b) and (c) for two different values of $N_{CZ}=1,\ 11$.
The solid lines in the figures are fits with $A_{1}=0.8$ (black), and $A_{11}=0.67$ (red) in both plots. Note that the decay of $A_{N_{CZ}}$ with the number of $CZ$ gates is due to decoherence. Finally, we remark here that this noise tailoring is not limited to injected coherent errors, and is also useful for tailoring low frequency charge noise and errors due to quasi-particle poisoning in our experimental device.

Next, we benchmark twirling of the measurement operations. We characterize the measurement error by extracting the confusion matrix $\mathcal{C}$. We prepare the two qubits in the computational basis states: $\{\ket{00},\ket{01},\ket{10},\ket{11}\}$, and immediately follow it by a measurement in the computational basis. The entries of the confusion matrix correspond to probabilities of measuring different computational basis states, where each row represents the prepared initial state, and each column corresponds to a measurement outcome. Since the dominant error mechanism during readout is qubit relaxation ($\ket{1}\rightarrow \ket{0}$), the confusion matrix is not symmetric as shown in \cref{fig:Twirled_Example} (d). This is consistent with the fact that relaxation error is not a Pauli channel. We verify that measurement twirling indeed symmetrizes the confusion matrix as shown \cref{fig:Twirled_Example}~(e). While this confusion matrix does not seperate out prepation and readout errors, the preparation errors from slighly elevated temperatures will also be largely symmetric.

\begin{figure*}
    \begin{minipage}[b]{\textwidth}
        \centering
        \begin{tikzpicture}
            \node at (-2.5,0.65) {\large Circuit $C\equiv$};
            \node (image) at (1.5,0.6) {
                \mbox{
                    \Qcircuit @C=.6em @R=.5em {
                        & \gate{S}   & \ctrl{1} & \gate{\sqrt{X}} & \ctrl{1}  &  \gate{\sqrt{X}} &  \ctrl{1} &  \gate{S}   &  \ctrl{1}  &\qw    \\
                        &  \gate{S}   & \ctrl{-1} &  \gate{\sqrt{X}}& \ctrl{-1} &  \gate{S}          &  \ctrl{-1} & \gate{S}  &  \ctrl{-1} & \qw \\ 
                    }
                }
            };
            \node at (6.75,0.65) {\large Input Pauli: $P_a=IY$, };
            \node at (10.75,0.65) {\large Output Pauli: $P_{a^\prime}=YY$};
            \node[ font=\bfseries, text=black] at (-4,1.2) {(a)};
            \node[ font=\bfseries, text=black] at (-4,-0.35) {(b)};
	\node[ font=\bfseries, text=black] at (-4,-2.7) {(c)};
            \node(image) at (4.5,-1.75) {
                    \scalebox{0.9}{
                        \mbox{
                                \Qcircuit @C=.7em @R=.5em {
                        & \\  
                        \lstick{\ket{0}} & \qw             & \qw       & \gate{S}  & \gate{\mathcal{E}_{S_{0}}}  & \ctrl{1}  & \multigate{1}{\mathcal{E}_{CZ}} & \gate{\sqrt{X}} & \gate{\mathcal{E}_{\sqrt{X}_{0}}} & \ctrl{1}  & \multigate{1}{\mathcal{E}_{CZ}} &  \gate{\sqrt{X}}  & \gate{\mathcal{E}_{\sqrt{X}_{0}}} &  \ctrl{1}  & \multigate{1}{\mathcal{E}_{CZ}} &  \gate{S}  & \gate{\mathcal{E}_{S_{0}}} & \ctrl{1}  & \multigate{1}{\mathcal{E}_{CZ}}  & \gate{\mathcal{E}_{SPAM_{0}}} &  \gate{\sqrt{X}} &   \qw  & \meter \\
                        \lstick{\ket{0}} & \qw & \gate{\sqrt{X}}  & \gate{S}  & \gate{\mathcal{E}_{S_{1}}}  & \ctrl{-1} & \ghost{\mathcal{E}_{CZ}}        & \gate{\sqrt{X}} & \gate{\mathcal{E}_{\sqrt{X}_{1}}} & \ctrl{-1} & \ghost{\mathcal{E}_{CZ}}        &  \gate{S}         & \gate{\mathcal{E}_{S_{1}}}        &  \ctrl{-1} & \ghost{\mathcal{E}_{CZ}}        &  \gate{S}  & \gate{\mathcal{E}_{S_{1}}} & \ctrl{-1} & \ghost{\mathcal{E}_{CZ}}         & \gate{\mathcal{E}_{SPAM_{1}}} &  \gate{\sqrt{X}} &   \qw  & \meter \gategroup{2}{4}{3}{20}{.7em}{--} \\ 
                                    &   &         &   \\
                                    &   &         &   
                                } 
                        }
                    }
                };
             \node at (-3.2,-0.75) {Input Pauli};
             \node at (4,-0.75) {Noisy Circuit};
             \node at (11.7,-0.75) {Output Pauli};
            \end{tikzpicture}

   \begin{tabular}{|c|}
        \hline
        Gate \\
        \hline
        Pauli \\
        \hline
        Element in A \\
        \hline
    \end{tabular}
    \begin{tabular}{|c|c|c|}
        \hline
        \multicolumn{3}{|c|}{$\sqrt{X}_{0}$} \\
        \hline
        X & Y & Z \\
        \hline
        1 & 1 & 0 \\
        \hline
    \end{tabular}
    \begin{tabular}{|c|c|c|}
        \hline
        \multicolumn{3}{|c|}{$\sqrt{X}_{1}$} \\
        \hline
        X & Y & Z \\
        \hline
        1 & 0 & 0 \\
        \hline
    \end{tabular}
    \begin{tabular}{|c|c|c|}
        \hline
        \multicolumn{3}{|c|}{$S_{0}$} \\
        \hline
        X & Y & Z \\
        \hline
        1 & 0 & 0 \\
        \hline
    \end{tabular}
    \begin{tabular}{|c|c|c|}
        \hline
        \multicolumn{3}{|c|}{$S_{1}$} \\
        \hline
        X & Y & Z \\
        \hline
        3 & 0 & 0 \\
        \hline
    \end{tabular}
    \begin{tabular}{|c|c|c|c|c|c|c|c|c|c|c|c|c|c|c|}
        \hline
        \multicolumn{15}{|c|}{$CZ$} \\
        \hline
        IX & IY & IZ & XI & XX & XY & XZ & YI & YX & YY & YZ & ZI & ZX & ZY & ZZ \\
        \hline
        0 & 0 & 0 & 0 & 0 & 1 & 0 & 0 & 0 & 2 & 0 & 0 & 1 & 0 & 0 \\
        \hline
    \end{tabular}
    \begin{tabular}{|c|c|c|}
        \hline
        \multicolumn{3}{|c|}{$SPAM_{0}$} \\
        \hline
        X & Y & Z \\
        \hline
        0 & 1 & 0 \\
        \hline
    \end{tabular}
    \begin{tabular}{|c|c|c|}
        \hline
        \multicolumn{3}{|c|}{$SPAM_{1}$} \\
        \hline
        X & Y & Z \\
        \hline
        0 & 1 & 0 \\
        \hline
    \end{tabular}

\end{minipage}

    \caption{Sample sequence representing a row of the design matrix. (a) Each row of the design matrix is specified by the tuple consisting of a circuit composed of gates drawn from the gate set, an input Pauli state $P_a$, and a corresponding output Pauli state $P_{a^{\prime}}$. (b) Full circuit corresponding to preparing the $-$ eigenstate of the input Pauli basis $P_a$, noisy circuit with noise elements as defined by the noise model, and output measurement basis specified by the output Pauli basis. (c) The elements of the design matrix $A$ for the row are specified by the conjugation of the input Pauli through the elementary gates composing the circuit, grouped by gate and Pauli here.}
    \label{fig:Circuit Sample}
\end{figure*}
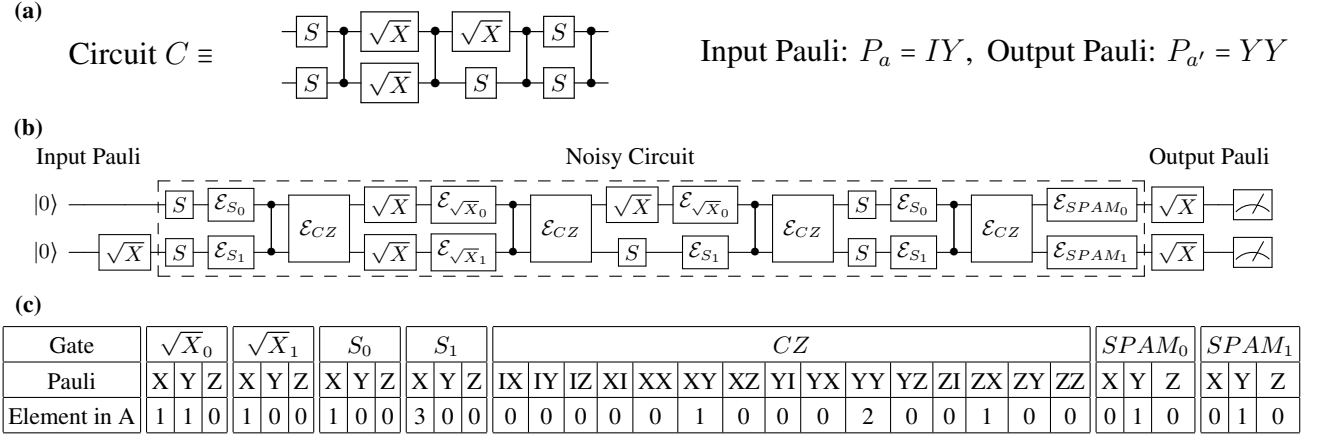

\subsection{\acrshort{aces} implementation details}
We discuss specific details concerning the \acrshort{aces} protocol. We cover error modeling, circuit generation, data collection, and fitting procedures.

\subsubsection{Error model}
We consider the following noisy Clifford gate-set for characterization using \acrshort{aces},
\begin{align}
    \mathcal{G}\equiv \{S_0,\sqrt{X_0},S_1,\sqrt{X_1},CZ\},
\end{align}
as discussed in  \cref{sec:expstep} where, we note that the $CZ$ gate corresponds to the composite (twirled) gate as defined in \cref{sec:twirling-implementation}. We consider a minimal model for the noise. We assume that the noise acts locally on the qubits that the gates are acting on; i.e., our error model does not capture any cross-talk. For example, the error model for a one-qubit gate acting on the first qubit consists of Pauli errors only on the first qubit. Similarly, the noise in the two-qubit $CZ$ gate is given by a two-qubit Pauli channel. Thus, our error model is  specified by $4\times 3+15$ parameters, corresponding to $27$ unknown error rates. For our fitting procedure, without any loss of generality, we choose to model the noise as being applied \emph{after} the application of the gate. 

The device has substantial preparation and readout noise which we incorporate into the noise model. However, it is well-known that state-preparation errors cannot be uniquely disambiguated from measurement errors~\cite{Flammia2020}. Thus, we model the effects of both errors with a single-qubit Pauli channel for each qubit, which results in two additional error channels described by $6$ more parameters.  We refer to these errors as \emph{SPAM} and include a more detailed discussion of the effects of noise in the preparation and measurement in the Appendix ( \cref{app:aces-theory}). In our noise model, the $SPAM$ errors are applied following the application of the circuit and prior to the measurement in the Pauli basis.

The full error model describing the experimental measurements consists of the following Pauli noise channels:
\begin{align}
    \mathbb{E}=\{\mathcal{E}_{\sqrt{X_0}},\mathcal{E}_{\sqrt{X}_1},\mathcal{E}_{S_0},\mathcal{E}_{S_1},\mathcal{E}_{CZ},\mathcal{E}_{SPAM_0},\mathcal{E}_{SPAM_1}\},
\end{align}
which is fully specified by 33 independent parameters. An example of application of the error model to a noiseless quantum circuit is provided in \cref{fig:Circuit Sample}(a) and (b).

\subsubsection{Constructing the design matrix}
\acrshort{aces} requires the construction of a full rank design matrix, $A$, from a collection of circuits that can be used to estimate the error rates associated with the individual operations. As discussed in \cref{sec:aces-review}, this design matrix can be constructed iteratively using tuples of Clifford circuits generated from the desired gate set, and input and output Paulis: $(C,P_a, P_{a^\prime})$. We iterate through randomly generated circuits containing alternating layers of single and two-qubit gates, along with different input Pauli states. In order to reduce experimental complexity, we prioritize input $Z$-Pauli states, and circuits with $3$-$5$ $CZ$ gates. The search of the circuit-Pauli combinations terminates when a full rank design matrix is achieved. Since the error model considered in the experiment has $33$ parameters, we anticipate needing at most 33 unique tuples to fit the parameters of the error model to the experimentally observed data. In \cref{fig:Circuit Sample} (c), we provide an example circuit and the corresponding row of the design matrix.

\subsubsection{Data collection and fitting}
 The experimental implementation involves the estimation of outcome probabilities for a full set of circuits (e.g., \cref{fig:Circuit Sample}(b)). These probabilities determine the circuit eigenvalues and are used to fit the Pauli error model. While only $33$ circuit eigenvalues are required to completely specify the error model, we collect multiple complete sets of \acrshort{aces} circuits to estimate the precision of the protocol. We measure $95$ different circuit eigenvalues which requires measurements from $192$ unique circuits; the difference comes from added requirements for the preparation of Pauli states. The outcome probabilities of the $192$ circuits are estimated by repeating it up to $N_{\rm shots}=10,000$ times. Pulling from these $95$ eigenvalues, we create $250$ unique full-rank design matrices $A$ that can be inverted to estimate the gate eigenvalues. Thus, each of the $250$ different combinations represents a complete set of ACES circuits that can be used to infer the error model. 

In our experimental device, the measurement error exhibits a temporal dependence that is not captured by the \acrshort{aces} error model, which assumes a \emph{time-independent} Pauli error model. In order to track and mitigate this drift error we interleave batches of \acrshort{aces} reconstruction circuits with measurement calibration benchmarking. We mitigate this drift error by performing \acrfull{mem}~\cite{Bravyi2021,Maciejewski2020}. The measurement error in each batch of circuits is mitigated by the corresponding confusion matrix; the procedure applies a pseudo-inverse of the transposed confusion matrix, $\left(\mathcal{C}^T\right)^{-1}$ (mitigation matrix), on the observed outcome probabilities to get the mitigated probabilities.

Next, we discuss the fitting procedure to extract the Pauli error model. Given a design matrix $A$ and experimentally measured vector of circuit eigenvalues $\boldsymbol{\Lambda}_C^{\rm exp}$ (see \cref{eq:aces-eq} for definitions), we estimate the error model using a bound-constrained optimizer (L-BFGS-B) with the following objective function,
\begin{align}
    \boldsymbol{p}^{\rm est}=\underset{\{\boldsymbol{p}\}}{\mathrm{argmin}}\left\|A\cdot\left(\log \left(\mathbb{H}(\boldsymbol{p})\right)\right)-\log(\boldsymbol{\Lambda}_C^{\rm exp}) \right\|_2.
\end{align}
Here, the bounds of the individual Pauli error rates for the non-identity Paulis were chosen to be $[0,0.2]$, and $\mathbb{H}(\cdots)$ represents the direct sum of Hadamard transforms applied to each one- and two-qubit Pauli error probabilities. Additionally, the error probabilities for each channel must separately sum to one; this constraint is incorporated in the Hadamard transformation by rewriting the error probability of the identity operation in terms of the other Paulis; e.g., $p_I=1-p_X-p_Y-p_Z$, for a one-qubit Pauli channel. We note here that,
in principle, the gate eigenvalues can be estimated efficiently via a standard least-squares solution to \cref{eq:aces-eq}. While the least-squares estimation leads to lower residuals for the circuit eigenvalues, the Pauli error models extracted from this fit often have unphysical, negative error probabilities. In such cases, one may obtain a \emph{valid} Pauli error model through a constrained optimization that minimizes the total-variation distance to the unphysical probability distributions subject to the constraint that the probabilities are positive. However, we find that the model obtained from this procedure typically has a large bias in the resulting residuals; see \cref{app:aces-theory} for example.

\section{Results}\label{sec:results}
We experimentally characterize the gate set on the two-qubit device with native and injected coherent errors. First, we perform a suite of standard characterization protocols based on \acrshort{rb} to extract average gate fidelities. Next, we extract a Pauli error model using the \acrshort{aces}. We compare the average gate fidelity predictions between the \acrshort{aces} and \acrshort{rb} approaches. 

\subsection{Randomized Benchmarking}\label{sec:rb}

\begin{figure*}
    \includegraphics[width=0.8\linewidth]{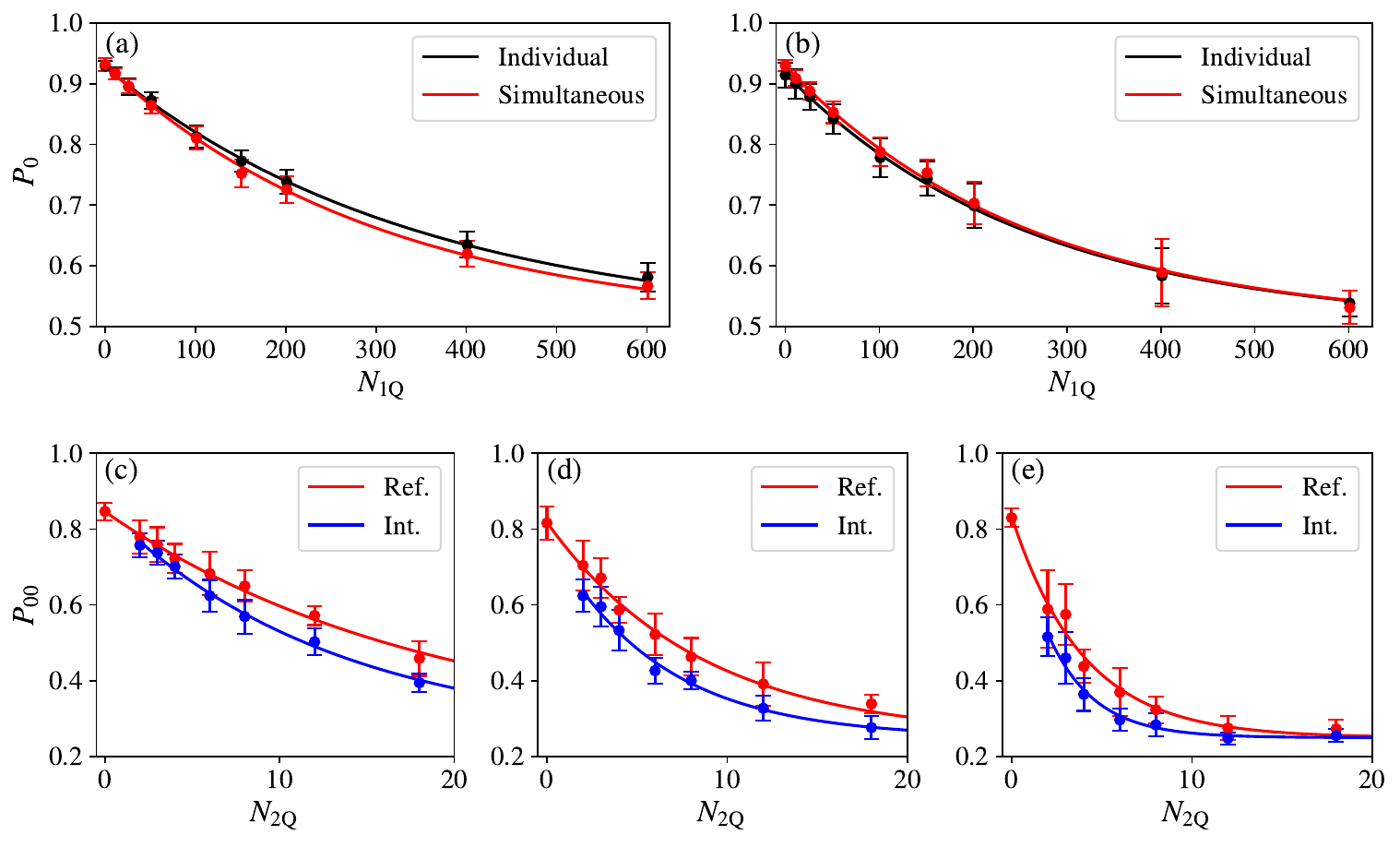}
    \caption{\acrshort{rb} for characterizing one and two-qubit gates. (Top row) Single-qubit, individual~(black) and simultaneous~(red) \acrshort{rb} decays of the outcome probability for each of the two qubits Q0~(a), and Q1~(b). (Bottom row) \acrshort{irb} for the twirled-$CZ$ gate with no error~(c), and added coherent error  $\Delta \theta = 0.06 \times 2\pi$~(d), $0.08 \times 2\pi$~(e). The two curves correspond to sequences with reference Clifford operations~(red), and sequence with the interleaved $CZ$ between the reference Cliffords (blue).  In all the figures, points denote experimental data and error bars represent the standard deviation across the different circuits at a given depth. Solid lines are fits to the curves. We report the \acrshort{rb} decay parameter in each figure:(a), and (b) $\alpha_{1,\rm ind(simul)}=0.9971(0.9967)$, $0.9962(0.9962)$; (c), (d), and (e)~$\alpha_{2,\rm ref(int)}=0.947 \pm 0.002 (0.926 \pm 0.003)$,~$0.889 \pm 0.005 (0.847 \pm 0.01)$,$0.778 \pm 0.014 (0.675 \pm 0.028)$.  The uncertainties reported here are associated with the fits and not the stability of the device.  The uncertainies with fitting $\alpha_{1}$ were negligible. }\label{fig:RB_data}
\end{figure*}

We characterize the one qubit gates in the device through individual and simultaneous \acrshort{rb}. The \acrshort{rb} protocol consists of a sequence of $N_{1{\rm Q}}$ random single-qubit Clifford gates applied to the qubits initialized to $\ket{0}$. When running the individual \acrshort{rb} of one qubit, the other qubit is kept idle. For simultaneous \acrshort{rb}, we apply the same number of Clifford gates on both qubits. Since \acrshort{rb} twirls over the Clifford group, the outcome probabilities are expected to decay with a single exponent. In the experimental implementation of each protocol, for a given $N_{1{\rm Q}}$, we run $20$ random Clifford sequences, where each Clifford is compiled into the native gate set. We fit the experimentally obtained outcome probabilities of each qubit to an exponential decay,
\begin{align}
    P_{0}= B_1\alpha_1^{N_{\rm 1Q}}+0.5,
\end{align} 
where $B_1$ and $\alpha$ are the fit parameters. The parameter $B_1$ accounts for any errors in state-preparation and measurement, and $\alpha_1$ is used to extract the average gate fidelity. We also note that the experimental implementation of the \acrshort{rb} protocols also included measurement twirling. As a result, the outcome probabilities asymptotes to $0.5$ as $N_{\rm 1Q}\rightarrow\infty$.
 For the one qubit Clifford gates, the average fidelity of a single Clifford gate can be estimated $\mathcal{F}_1=1-\frac{1}{2}(1-\alpha_1)$. In \cref{fig:RB_data}~(a) and \cref{fig:RB_data}~(b), we compare both \acrshort{rb} protocols on the two qubits. Fitting the individual \acrshort{rb} data, we find that the single-qubit average gate fidelities are $0.9986$ and $0.9981$ for ${\rm Q0}$ and ${\rm Q1}$ respectively. From simultaneous \acrshort{rb} we find average gate fidelities of 0.9984 (${\rm Q0}$) and 0.9981 (${\rm Q1}$). For both qubits there is no significant difference between individual and simultaneous \acrshort{rb}, indicating minimal cross-talk in the implementation of one qubit gates. 

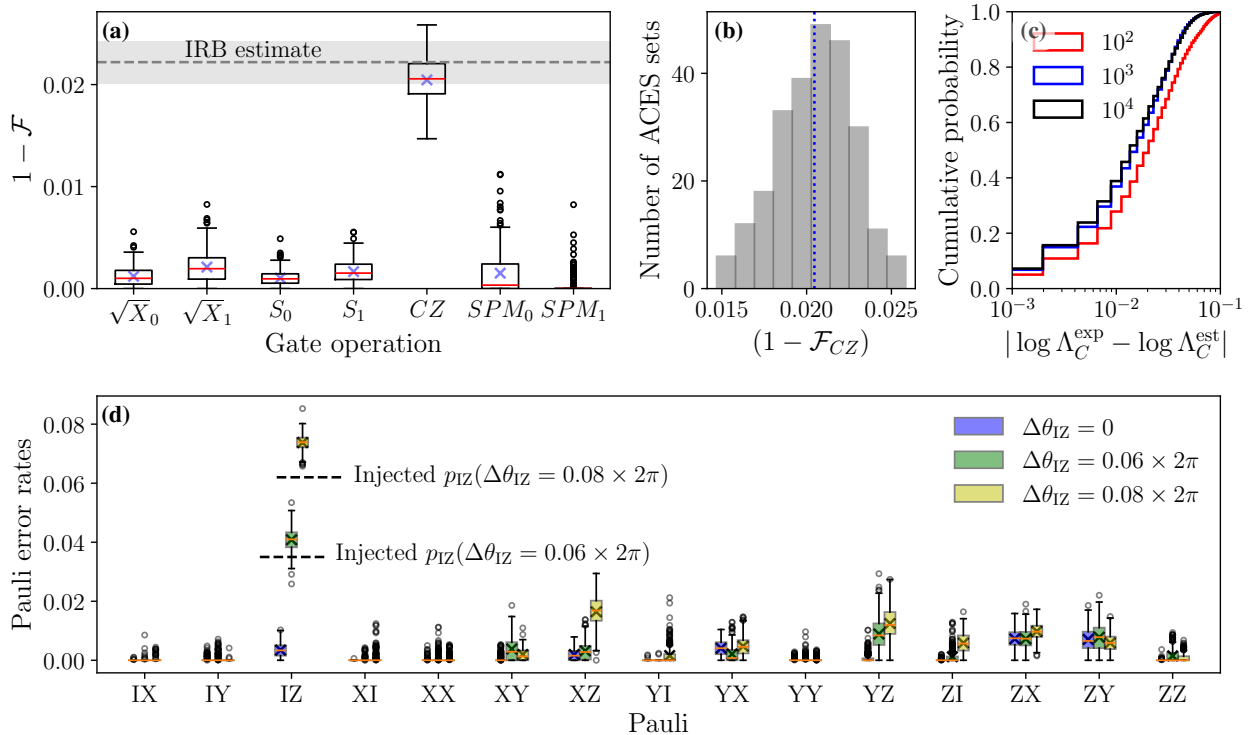
\begin{figure*}
        \centering
        \begin{tikzpicture}
            \node(image) at (0,0) {
                \includegraphics[scale=.65]{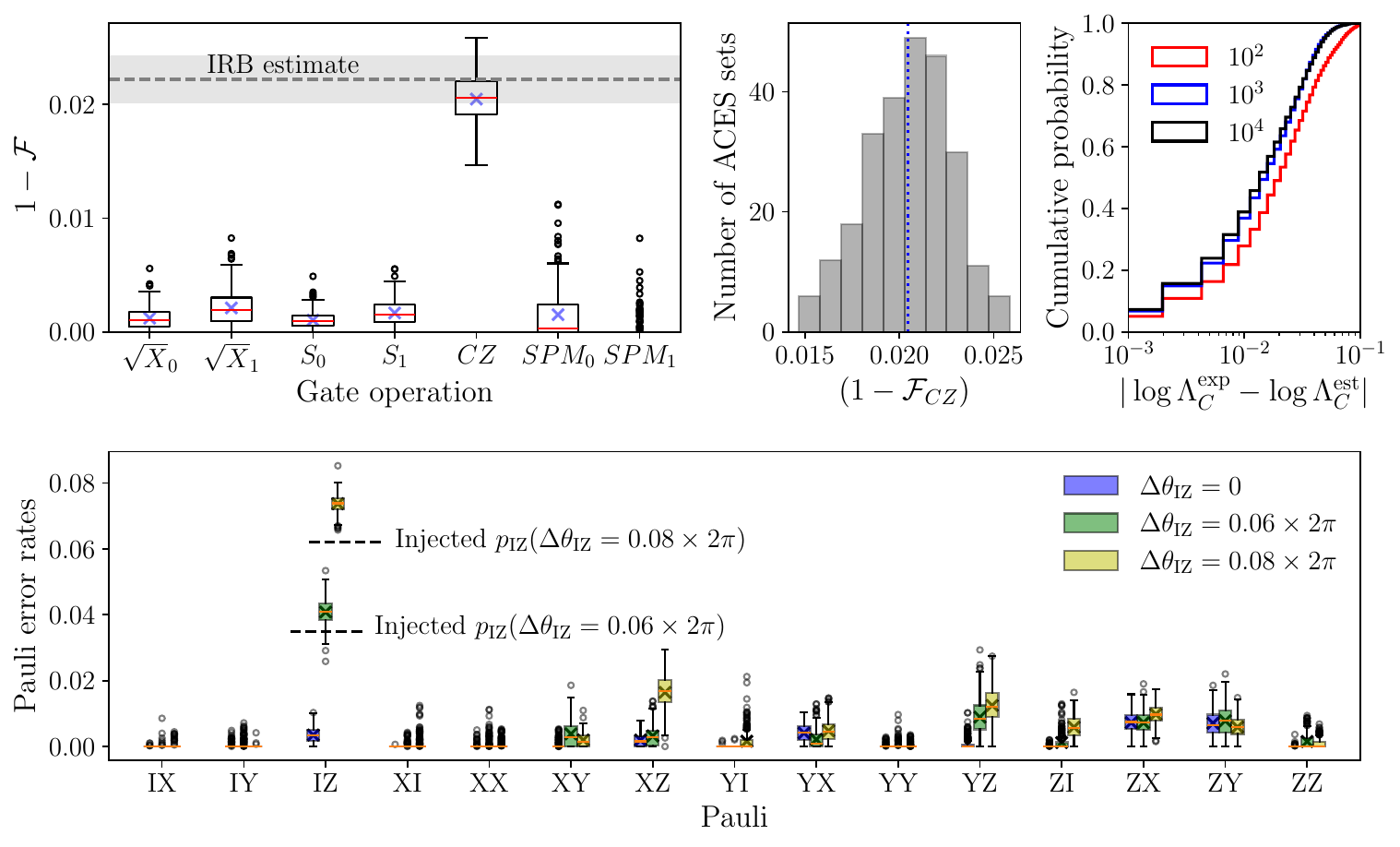}
            };
            \node[ font=\bfseries, text=black] at (-6.8,4.5) {(a)};
            \node[ font=\bfseries, text=black] at (1.35,4.5) {(b)};
            \node[ font=\bfseries, text=black,fill=white,fill opacity=0.7,rounded corners] at (5.4 ,4.5) {(c)};
            \node[ font=\bfseries, text=black] at (-6.8,-0.6) {(d)};
        \end{tikzpicture}
 \caption{\acrshort{aces} for characterizing Pauli noise on a two-qubit device. (a)~Average gate infidelities ($1-\mathcal{F}$) for all one and two-qubit gates in the gate set. The error models are constructed from $250$ distinct design matrices, with a standard box and whisker representation of the data with the blue cross representing the mean. The predicted fidelity from \acrshort{irb} is shown as a dashed line. (b)~The histogram of the average fidelity of the $CZ$ gate as computed from the $250$ \acrshort{aces} fits with the mean shown as the blue dashed line. (c)~Cumulative probability of a given circuit residual across all $250$ fits for the circuit eigenvalues between experimentally observed ($\Lambda_C^{\rm est}$) and that predicted by the error model $\Lambda_C^{\rm est}$. (d) All Pauli error rates for the $CZ$ gate for different strengths of injected coherent error $\Delta\theta=0$, $0.06\times 2\pi$, and $0.08\times 2\pi$.} \label{fig:aces}
\end{figure*}

Next we characterize the twirled-$CZ$ gate implementation (e.g., see \cref{fig:PauliTwirlError}) using the \acrshort{irb} protocol~\cite{Magesan_2012}. \acrshort{irb} estimates the average fidelity of a specific gate by comparing \acrshort{rb} decays with and without the gate interleaved into each circuit. For a two-qubit gate, the protocol compares two-qubit \acrshort{rb} decay with a two-qubit \acrshort{rb} protocol that has an interleaved $CZ$ gate. Explicitly, the \acrshort{rb} decays are fit to the following parametric equation: $P_{00}^{\rm int(ref)}=B_2(\alpha_{2,{\rm int(ref)}})^{N_{\rm 2Q}}+0.25$. The two-qubit gate fidelity is then estimated from the two exponential decays via,
\begin{align}
    \mathcal{F}_{CZ}=1-\frac{3}{4}\left(1-\frac{\alpha_{2,{\rm int}}}{\alpha_{2,{\rm ref}}}\right).
\end{align}
The experimental implementation of \acrshort{irb} is shown in \cref{fig:RB_data}~(c)-(e). We estimate the fidelity for the twirled-$CZ$ natively and with injected coherent errors as a virtual $R_z(\Delta\theta)$ gate; see \cref{fig:PauliTwirlError}.  While \cref{fig:RB_data}~(c) has no injected error on the $CZ$ gate, \cref{fig:RB_data}~(d) and \cref{fig:RB_data}~(e) correspond to $\Delta \theta = 0.06 \times 2\pi$ and   $0.08 \times 2\pi$, respectively. In all three cases, we run $10$ circuits per depth $N_{\rm 2Q}$ and $256$ shots for each circuit, and we obtain exponential decays with~(blue) and without~(red) an interleaved $CZ$ gate. We note that this is the case even though the injected error is coherent since the $CZ$ gate is Pauli-twirled at every shot. Thus, the twirling tailors the coherent error into a stochastic Pauli channel. Finally, we see that the exponential decays also saturate at $0.25$, consistent with measurements on both qubits being twirled. Fitting the exponential decay we estimate the interleaved twirled CZ average gate fidelity without any injected error as $\mathcal{F}_{CZ} = 0.983 \pm 0.003$. This decreases to $\mathcal{F}_{CZ} = 0.964 \pm 0.009 $ and $0.90 \pm 0.03$ for injected coherent rotations of $0.06 \times 2\pi$ and $0.08 \times 2\pi$ respectively. We note here that while the \acrshort{rb} results discussed in the section are representative of the general qualitative behavior of the device, the measured two-qubit gate fidelities exhibit temporal variations (at the scale of minutes) by up to $0.01$; see \cref{app:devicedetails} for a discussion on this time-dependence. 

We make a couple of remarks about the shortcomings of the standard \acrshort{rb} approach to characterizing the experimental gate-set. For one qubit \acrshort{rb}, the protocol estimates the average gate fidelity of a one-qubit Clifford gate, but does not give estimates of fidelity for each gate separately. While \acrshort{irb} can be used to estimate a gate fidelity of the two-qubit gate, it returns no information about the error mechanisms responsible. Furthermore, it is well understood that the accuracy of \acrshort{irb} can strongly depend on the fidelity of the reference sequence~\cite{Magesan_2012}. As we discuss next, \acrshort{aces} can be used to address some of these shortcomings.

\subsection{\acrshort{aces}}

\cref{fig:aces}~(a) shows the average gate infidelity for all the gates as obtained from the $250$ different design matrices constructed from the $95$ circuit eigenvalues using the \acrshort{aces} protocol. We note that the one-qubit gates have infidelities that are an order of magnitude smaller than the $CZ$ gates, in line with what is expected from the analysis in \cref{sec:rb}. In \cref{fig:aces}~(b), we examine the variation in the estimated fidelity across the different complete sets. The histogram of the estimates of the gate fidelity is uni-modal, and the mean is a good statistic for estimating the average fidelity. Additionally, in order to cross-check the estimated gate fidelities from \acrshort{aces}, we use \acrshort{irb} to estimate the $CZ$ gate fidelity directly before a complete set of \acrshort{aces} circuits were run. The average gate fidelity was estimated to be $\mathcal{F}_2 = 0.978 \pm 0.002$, shown as the dashed line and solid bar in \cref{fig:aces}~(a), and in agreement with the \acrshort{aces} estimate. We note here that the low estimate for error for $SPAM$ on both qubits is a reflection of the fits being performed on data after \acrshort{mem} has been applied.

We directly compare the experimentally measured circuit eigenvalues to that predicted by the \acrshort{aces} error model. \cref{fig:aces}~(c) shows the residuals for the circuit eigenvalues ($33$ per realization) from the $250$ different fits. The plot shows the cumulative probability of a given residual for increasing number shots. We see clear improvement in our fit by increasing the number of shots from $10^2$ (red) to $10^3$ (blue). However, the gains are only marginal going from $10^3$ to $10^4$. This leads us to believe that we have a sufficient number of shots in our circuit eigenvalue estimation step and the fitting procedure is not limited by shot noise.

Next we examine the full Pauli error channel for the twirled-$CZ$ gate. To evaluate the accuracy of the fitting procedure, we compare the estimated error model in the presence of intentionally injected coherent errors on the qubit Q1, as described in Fig. ~\ref{fig:PauliTwirlError}. Twirling tailors a coherent  error on Q1 into a Pauli error. Thus, this noise injection should be flagged as an increase in the $p_{IZ}$ error rate for the two-qubit Pauli channel. The distribution of individual Pauli error rate estimates for the three different levels of injected error are shown in \cref{fig:aces}~(d). The black dashed lines indicate the ideal error rate, $p_{IZ}$ based only on twirling the injected coherent error. Clearly, \acrshort{aces} identifies the increase in  $p_{IZ}$ with increasing magnitude coherent error. We note that the estimated $p_{IZ}$ is always greater than the ideal injected error. This difference can be traced to a contribution from the noise channel of the bare $CZ$ gate in the absence of any injected noise. Another source of the discrepancy is likely drift in the device, which may lead to different error rates for the bare gates at different points in time. We conclude that \acrshort{aces} is able to accurately identify the error source corresponding to the intentional errors and roughly estimate the amount of error on each Pauli term.

 \section{Discussion and Outlook}\label{sec:discussion}

In this paper, we have examined the \acrshort{aces} protocol for characterizing noise in one- and two-qubit gates in a two-qubit device. We find that \acrshort{aces} is able to accurately estimate Pauli error rates for the gate set. A key requirement for accurate characterization is that the errors in gates are well described by Pauli channels. This is achieved in experiment through twirling that is implemented on a shot-by-shot basis. The main advantage of \acrshort{aces}, when compared to traditional \acrshort{rb} based approaches, is that the protocol is fast, scalable to many qubits, and recovers a detailed Pauli error model for all the gates simultaneously. We posit that \acrshort{aces} as a \acrshort{qcvv} protocol is useful for tracking slow drifts in gate performance.

One of the main drawbacks of \acrshort{aces} is the tomographic nature of the error reconstruction; i.e., the accuracy of the reconstruction is affected by imperfect preparation and measurement of Pauli states. While we do account for some of these errors through the $SPAM$ error channel, this model cannot capture non-Pauli errors, as well as errors during both state-preparation and measurement; we investigate this numerically in \cref{app:aces-theory}. This issue is reflected in the large value of residuals obtained from the \acrshort{aces} error model. We conjecture that a significant source of error in the experiment is in the initial state-preparation. We estimate the steady state qubit temperature of 46 mK which corresponds to an excited state population of 2.1\% and 2.7\%  for Q0 and Q1, respectively~\cite{Geerlings2013}. 
Another limitation of \acrshort{aces} is in capturing any temporal variations of error rates during the execution of the protocol. These errors could be a slowly changing coherent error in the gates, or low-frequency $1/f$ noise and are known to be prevalent in superconducting qubits~\cite{Oliver2020}. We discuss temporal variations in the device in \cref{app:devicedetails}. Other sources of error not captured with \acrshort{aces} include a hot tunable coupler and leakage from non-adiabatic flux pulses.

There are several directions to pursue further development of \acrshort{aces}. Recent work has extended Pauli noise learning protocols for learnable degrees of freedom to gate sets~\cite{Chen2025}. An interesting area of exploration is how these protocols can be adapted to characterize temporally correlated noise. Another topic of research is extending these protocols to characterizing mid-circuit measurements, a key component of quantum error correction protocols. Another fruitful avenue to develop is a scalable protocol for benchmarking leakage errors which can be quite detrimental for quantum error correction.

\begin{acknowledgements}
The qubit devices used in these study were fabricated at MIT Lincoln Laboratory under the program Superconducting Qubits at Lincoln Laboratory (SQUILL).
\end{acknowledgements}
\appendix

\section{Temporal drift in device}\label{app:devicedetails}

\begin{figure}[h]
    \includegraphics[width=\linewidth]{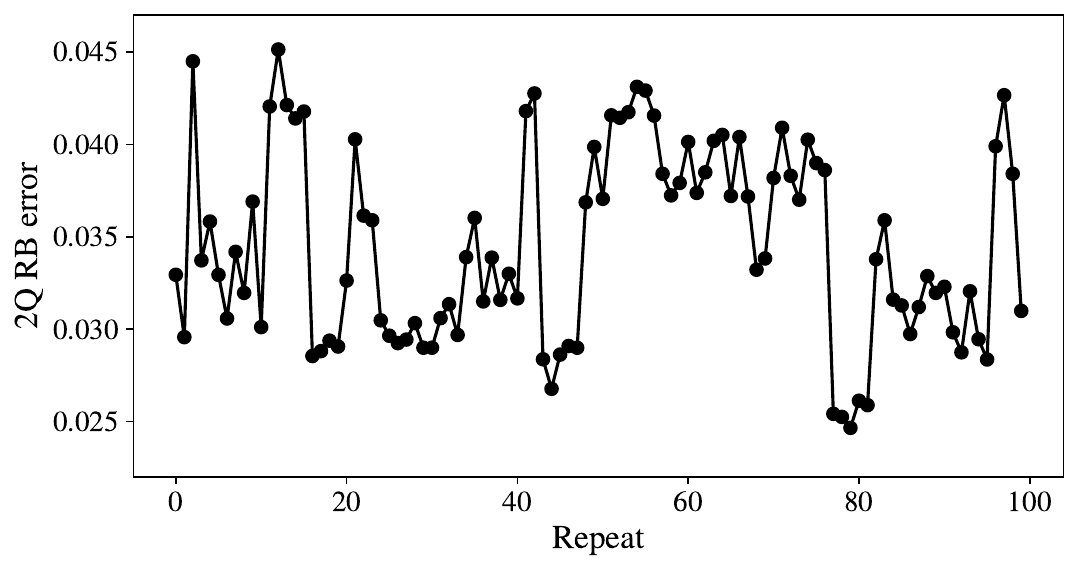}
    \caption{Average two-qubit gate error as extracted from consecutive two-qubit \acrshort{rb} experiments. }\label{fig:RBdrift}
\end{figure}

As mentioned in the main text, the device examined in this paper exhibits temporal variations on the scale of minutes. We present data from two-qubit \acrshort{rb} analysis that was run consecutively $100$ times over approximately a period of $600$ minutes. As shown in \cref{fig:RBdrift}, the average Clifford error rate extracted from the protocol exhibits sharp fluctuations. These fluctuations appear to be correlated with fluctuations in the offset charge; a consequence of the fabricated device having lower $E_J/E_C$ ($\approx 35$) than is typical for transmons.

\section{Performance analysis of \acrshort{aces}}\label{app:aces-theory}
In this appendix, we provide additional analysis of the performance of \acrshort{aces}. We discuss the quality of the fitting procedure in the presence of errors that are not captured by the error model. Next, we compare optimizers used for the fits: the bound-constrained optimizer used in the main text is compared to a two-step optimization process that involves a least-squares fit of the gate eigenvalues followed by a convex, constrained optimization to estimate the error probabilities. 

\subsection{Out-of-model errors}
The performance of \acrshort{aces} is benchmarked by analyzing the residuals for the fits to data that is numerically generated from different models. We show that the \acrshort{aces} fits are limited by sampling noise when the errors generating the data are well described by the \acrshort{aces} error model. However, in the presence of noise that is not perfectly captured by the error model (e.g, initialization errors), the residuals are limited by these noise processes.  

\begin{table}[b]
    \begin{tabular}{|c|c|c|c|c|c|c|c|}
        \hline
        \multicolumn{2}{|c|}{$\sqrt{X}_{0}$} & \multicolumn{2}{|c|}{$\sqrt{X}_{1}$} & \multicolumn{2}{|c|}{$S_{0}$} & \multicolumn{2}{|c|}{$S_{1}$} \\
        \hline
        $p_I$ & $0.998$   & $p_I$ & $0.997$   & $p_I$ & $0.999$   & $p_I$ & $0.998$ \\
        $p_X$ & $1.28e-5$ & $p_X$ & $1.52e-4$ & $p_X$ & $1.12e-3$ & $p_X$ & $1.07e-3$ \\
        $p_Y$ & $1.65e-3$ & $p_Y$ & $2.00e-3$ & $p_Y$ & $3.15e-4$ & $p_Y$ & $4.29e-4$ \\
        $p_Z$ & $1.40e-4$ & $p_Z$ & $1.02e-3$ & $p_Z$ & $1.09e-4$ & $p_Z$ & $9.93e-4$ \\
        \hline
        \multicolumn{8}{|c|}{$CZ$} \\
        \hline
        $p_{II}$ & $0.974$   & $p_{IX}$ & $1.12e-5$ & $p_{IY}$ & $1.90e-4$ & $p_{IZ}$ & $3.45e-3$ \\
        $p_{XI}$ & $2.52e-6$ & $p_{XX}$ & $1.23e-4$ & $p_{XY}$ & $4.53e-4$ & $p_{XZ}$ & $1.95e-3$ \\
        $p_{YI}$ & $1.66e-5$ & $p_{YX}$ & $4.08e-3$ & $p_{YY}$ & $1.05e-4$ & $p_{YZ}$ & $7.74e-4$ \\
        $p_{ZI}$ & $7.73e-5$ & $p_{ZX}$ & $7.24e-3$ & $p_{ZY}$ & $7.06e-3$ & $p_{ZZ}$ & $3.70e-5$ \\
        \hline
    \end{tabular}
    \caption{Error probabilities for each gate operation used in gate-error model. These probabilities were calculated from experimental data and the resulting noise model is used for our simulations involving out-of-model errors.}
    \label{table:ExpErrorProbs}
\end{table}

We generate the data for \acrshort{aces} using four different error models:
\begin{enumerate}
    \item Model I~(\emph{ideal prep \& ideal measure}): A depolarizing error model for all gates involved in the circuit, with perfect Pauli state preparation and measurement.
    \item Model II~(\emph{ideal prep \& non-ideal measure}): Same gate-error model as Model I, but with a symmetric readout error as modeled by a bit-flip error on each qubit.
    \item Model III~(\emph{non-ideal prep \& ideal measure}): Same gate-error model as Model I, but with a bit-flip error associated with initialization of the qubits to $\ket{0}$.
    \item Model IV~(\emph{non-ideal prep \&  non-ideal measure}): Same gate-error model as Model I with readout error as in Model II, and prep error as in Model III.
\end{enumerate} 
The gate-error model is chosen to be an asymmetric depolarizing error model based on estimates from the experiments in the main-text. The depolarizing error probabilities are shown in \cref{table:ExpErrorProbs}.

\begin{figure}[htbt]
    \includegraphics[scale=.6]{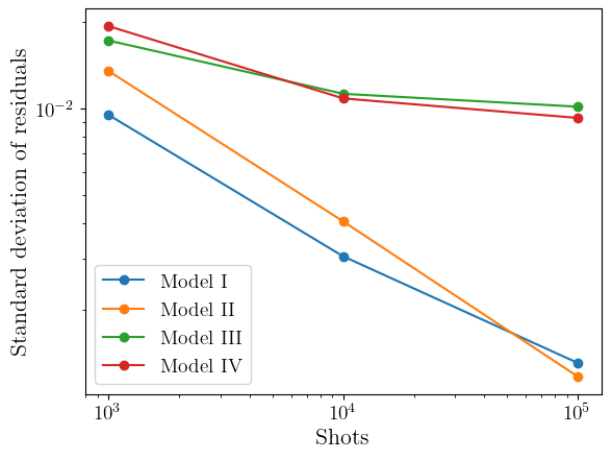}
    \caption{Circuit-eigenvalue residuals for numerically generated data from four different models described in the text.} \label{fig:app-aces_theory_noise}
\end{figure}

Readout error is modeled as a bit-flip error with error probabilities, $\{p_{M_0}=0.02,p_{M_1}=0.05\}$.  Initialization error is modeled as a bit-flip error with error probabilities, $\{p_{R_0}=0.019,p_{R_1}=0.024\}$. The initialization error is chosen based on independent temperature measurements performed on the device.

In \cref{fig:app-aces_theory_noise} we summarize the dependence of circuit-eigenvalue residuals, as estimated by the standard-deviation, on the sampling number. It is clear that for Models I and II, the circuit-eigenvalue residuals reduces with the number of samples. This indicates that the \acrshort{aces} error model is accurately fitting the generated data. This is to be expected, as the default noise model considered in \acrshort{aces} fits this picture. However, once we add a bit-flip error to the initialization (as is the case in Models III and IV), the circuit-eigenvalue residuals plateau with increasing sampling number. This indicates that the residuals are limited by errors not captured by the \acrshort{aces} error model.

\subsection{Optimizer comparisons}
\begin{figure}[h]
    \includegraphics[scale=.6]{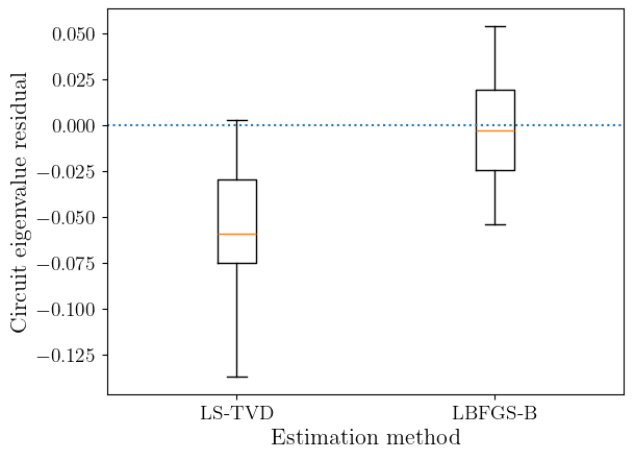}
    \caption{Comparison of circuit eigenvalue residuals from least-squares+tvd minimization, and L-BFGS-B solvers on one illustrative set of \acrshort{aces} circuits.}
    \label{fig:app-aces_theory_optimizers}
\end{figure}

In the main text, we described the estimation procedure by recasting the \acrshort{aces} equations in terms of the Pauli error probabilities. Here, we compare its performance with the least-squares based approach. Typically, the least-squares solution to \cref{eq:aces-eq} has lower circuit eigenvalue residuals; however, the gate-eigenvalues obtained as a solution often lead to invalid Pauli error channels (as obtained through a Hadamard transform) due to the presence of negative probabilities. One may fix this issue by identifying probability distributions closest (in total variation distance) to the Pauli error probabilities obtained from the gate eigenvalues using a constrained, convex optimization. In general, we find that the resulting Pauli error model tends to generate a large bias in the circuit eigenvalue residuals. This bias in the residuals is shown in an example fit in \cref{fig:app-aces_theory_optimizers}.
\bibliographystyle{ieeetr}
\bibliography{refs}
\end{document}